\documentstyle[prb,aps,epsf]{revtex}
\begin{document}
\advance\textheight by 0.5in
\advance\topmargin by -0.25in
\draft

\twocolumn[\hsize\textwidth\columnwidth\hsize\csname@twocolumnfalse%
\endcsname

{\hfill\normalsize NSF--ITP--96--148, cond-mat/9611126\medskip\\}

\title{Correlation Effects in Carbon Nanotubes}

\author{Leon Balents and Matthew P. A. Fisher}
\address{Institute for Theoretical Physics, University of California,
Santa Barbara, CA 93106--4030}

\date{\today}

\maketitle

\begin{abstract}
  We consider the effects of Coulomb interactions on single-wall
  carbon nanotubes using an on-site Hubbard interaction, $u$.  For the
  (N,N) armchair tubes the low energy theory is shown to be {\it
    identical} to a 2-chain Hubbard model at {\it half-filling}, with
  an effective interaction $u_N = u/N$.  Umklapp scattering leads to
  gaps in the spectrum of charge and spin excitations which are
  exponentially small for large $N$.  Above the gaps the intrinsic
  nanotube resistivity due to these scattering processes is linear in
  temperature, as observed experimentally.  The presence of ``d-wave"
  superconductivity in the 2-chain Hubbard model away from
  half-filling suggests that {\sl doped} armchair nanotubes might
  exhibit superconductity with a purely electronic mechanism.

\end{abstract}
\vskip -0.5 truein
]

Carbon nanotubes constitute a novel class of quasi one-dimensional
(1d) materials which offer the potential for both new physics and
technology.\cite{Review}\  Although built only with carbon atoms, they
can be grown in a tremendous variety of shapes and sizes.  The
simplest single-wall tube consists of a single graphite sheet which is
curved into a long cylinder, with a diameter which can be smaller than
1nm.  Several groups\cite{Single}\ have succeeded in measuring the
resistance of a {\it single} multi-wall nanotube, composed of several
concentric cylinders.  Crystalline ``ropes" consisting of a triangular
packing of (nominally) identical single-wall tubes are also very
promising, exhibiting signatures of metallic transport.\cite{Ropes}\

Generally, single wall tubes can be characterized by two integers,
$(N,M)$, which specify the super-lattice translation vector which wraps
around the waist of the cylinder.  Current
theories\cite{Theory1,Theory2}\ consist of band structure calculations
and predict a rich variety of behavior, ranging from metallic
``armchair" tubes with $(N,N)$ to insulating ``zig-zag" tubes with
$(N,0)$.  For very small nanotubes, however, electron correlation
effects should become important, as in other 1d systems.\cite{1d}\  In
this paper we study these effects using a tight-binding description
(which correctly reproduces the band-structure calculations)
supplemented by an on-site Hubbard interaction $u$.  For the $(N,N)$
armchair tubes we show that the effective description at low energies
is {\it identical} to a 2-chain Hubbard model at {\it half-filling}
with an effective interaction strength, $u_N=u/N$.  Since this
effective interaction is weak for $N \sim 10$, its effects can be
treated perturbatively.  Particularly important are electronic Umklapp
scattering processes, present at half-filling.  These are predicted to
open a small charge and spin-gap, changing the behavior from metallic
to insulating at low enough temperatures.  Similar conclusions have
been reached independently in very recent work by Krotov, Lee and
Louie.\cite{scoop}\  At temperatures above the charge gap, a simple
weak-coupling analysis of these interactions gives a resistivity which
varies linearly with temperature, which may explain the observed
behavior\cite{Ropes}\ in single carbon ``ropes".  Furthermore, doping
an armchair nanotube is equivalent to moving away from 1/2-filling in
the 2-chain Hubbard model.  This problem has been extensively
studied,\cite{Hubbard}\ and exhibits superconducting behavior at low
temperatures, with a ``d-wave" symmetry.  This suggests a possible
electronic mechanism for superconductivity in doped nanotubes.

\begin{figure}[hbt]
\epsfxsize=\columnwidth\epsfbox{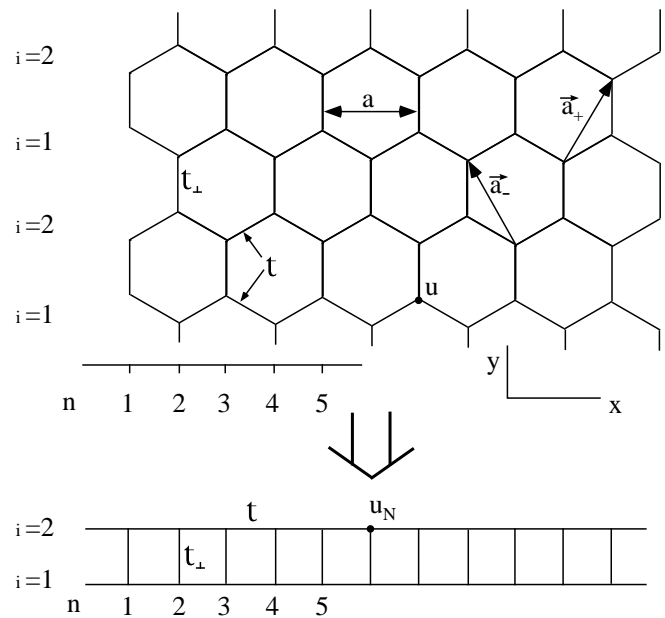}
\vspace{15pt}
\caption{Illustration of the graphite lattice, with labeling and
  periodic boundary conditions for an armchair tube.}
\label{fig1}
\end{figure}

Following various authors,\cite{Theory1}\ we first consider a single
sheet of graphite, composed of carbon atoms arranged on the sites of a
honeycomb lattice.  The underlying Bravais lattice is triangular, with
two sites per unit cell.  The two primitive Bravais lattice vectors
are $\vec a_{\pm} = (a/2)(\pm 1, \sqrt{3})$, where $a= \sqrt{3} d$,
with $d$ the near-neighbor carbon separation.  Of the four outer shell
electrons of each atom, three form the $sp_2$ bonds of the lattice,
while the fourth can tunnel between neighboring $p_z$ orbitals.  A
simple description, which correctly accounts for the semi-metallic
behavior of graphite, consists of a tight binding model with one $p_z$
orbital per carbon, and a tunneling matrix element $t$ between
neighboring atoms.  The Bloch states for this tight-binding model form
two bands, with energies $E_\pm(\vec k) = \pm [ |\xi(\vec
k)|^2]^{1/2}$ where
\begin{equation}
  \xi(\vec k) = 2t \cos(k_xa/2) e^{ik_ya/2\sqrt{3}} + t_{\perp}
  e^{ik_ya/\sqrt{3}}   ,
\end{equation}
and $\vec k$ is the crystal momentum.  Here we have allowed for a
different hopping strength, $t_\perp$, in the $y$-direction (see
Fig.~1).  With one electron per carbon atom, the Fermi energy is at
$E=0$, with the lower band full and the upper empty.  The striking
feature of this band structure is that there are two isolated points in
the first Brillouin zone, denoted $\vec K_{\pm}$, where the bands
touch $E=0$, and there are gapless excitations.  In the vicinity of
these ``Dirac" points, for $\vec q = \vec k - \vec K_{\pm}$ small, the
dispersion is relativistic, with $E(\vec q) = v |\vec q|$ and
$v=(\sqrt{3}a/2)t$ (for $t_\perp =t$).  When $t=t_\perp$, the gapless
points occur at $\vec K_{\pm} = (\pm 4\pi/3a,0)$, but are shifted
along the $k_x$ axis for $t \ne t_\perp$ (see Fig.~2).

\begin{figure}[hbt]
\epsfxsize=\columnwidth\epsfbox{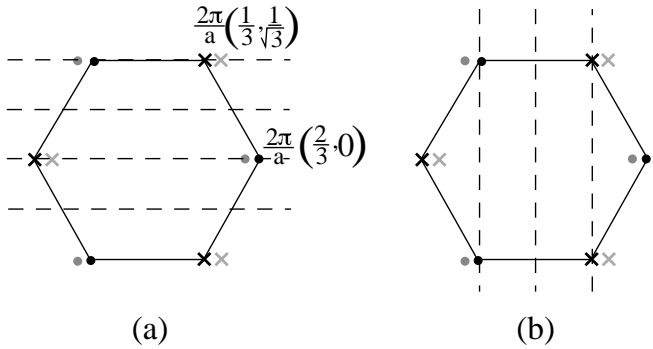}
\vspace{15pt}
\caption{Dirac points in the Brillouin zone.  Dark circles and crosses
  indicate the locations of the gapless points for $t_\perp = t$,
  while gray symbols schematically indicate the shifted positions for
  $t_\perp < t$.  Dashed lines cut the zone at a discrete set of
  allowed transverse momenta in (a) the armchair tube and (b) the
  zig-zag tube (here with $N=3$).}
\label{fig2}
\end{figure}

Single wall nantotubes consist of rolling the honeycomb sheet of
carbon atoms into a cylinder.  Each tube is characterized by two
integers\cite{Theory1}\ $(N,M)$, which specify the super-lattice
translation vector $T_{(N,M)} = N \vec a_+ + M \vec a_-$, which wraps
around the waist of the cylinder.  The crystal momentum transverse to
the axis of the cylinder is then quantized.  Band structure
predicts\cite{Theory1}\ metallic behavior whenever the gapless points
in the Brillouin zone lie on the allowed transverse quantized
wavevectors.  For the armchair nanotubes with $(N,N)$ this is
illustrated in Fig.~2a, where the allowed values of $k_y$ are shown as
dashed lines for $N=4$.  Since gapless modes are present at $k_y=0$,
band structure predicts metallic behavior for armchair tubes,
independent of $N$.  Due to curvature effects,\cite{Theory2}\ the
hopping matrix elements along ($t$) and around ($t_\perp$) the
nanotube will differ slightly, by an amount of order $1/N^2$.  This
shifts the Dirac points at $\vec K_{\pm}$ along $k_x$, but leaves the
armchair tube gapless (metallic).  For the $(N,-N)$ zig-zag tube
(equivalent to the $(N,0)$ tube) with non-integer $N/3$, the gapless
points do not coincide with quantized transverse momenta, so that
insulating behavior is predicted with a gap varying as $1/N$.  For
integer $N/3$ the Dirac points for $t=t_\perp$ are at quantized
transverse momenta, but are shifted away slightly, of order $1/N^2$,
due to curvature effects\cite{Theory2} ($t \ne t_\perp$).  Thus
band structure predicts semi-metallic behavior for integer $N/3$ zigzag
tubes (Fig.~2b).

\begin{figure}[hbt]
\epsfxsize=\columnwidth\epsfbox{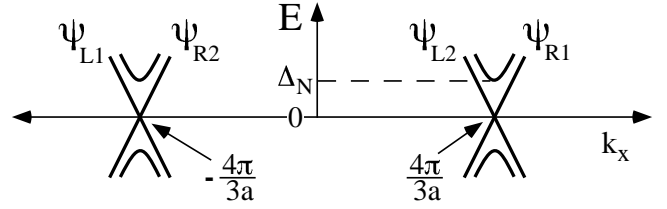}
\vspace{15pt}
\caption{One-dimensional spectrum near Dirac points.}
\label{fig3}
\end{figure}

For the armchair tubes the low energy modes
occur near the two gapless points, at $k_y=0$.  The
1d dispersion away from these two points is shown in
Fig.~3.  In addition there are gapped modes at $k_y \ne 0$, with an
energy of order $\Delta_N = v(2\pi /\sqrt{3} N)^2$.  Below $\Delta_N$,
the mode structure is equivalent to a 1d 2-band model,
independent of the nanotube size $N$.
 
Coulomb interactions can be incorporated into the nanotube tight
binding model, and will introduce interactions into the effective 1d
model.  To be concrete we focus on an on-site Hubbard interaction, $u$,
so that the full Hamiltonian becomes,
\begin{equation}
  H = -\sum_{<{\bf r}{\bf r}^\prime>} t_{{\bf r}{\bf r}^\prime}
  c^\dagger_{\alpha}({\bf r}) 
  c_{\alpha}({\bf r}^\prime) + u \sum_{\bf r} n_{\uparrow}({\bf r})
  n_{\downarrow}({\bf r})  ,
\end{equation}
where the first sum is over spin states ($\alpha = \uparrow,
\downarrow$) and near neighbor sites of the honeycomb lattice, and
$n_{\alpha}({\bf r}) = c^\dagger_{\alpha}({\bf r}) c_{\alpha}({\bf
  r})$.  We now show that for $(N,N)$ armchair tubes the effective
interacting 1d model is {\it identical} to a 2-chain Hubbard model
with an interaction strength, $u_N = u/N$.  To do so, we choose a
particular basis of states spanning the space of low energy states
with $k_y=0$:
\begin{eqnarray}
  \phi_{n1}(x,y) & = & \cases{ 
    N^{-1/2}\delta_{x,na_0} \delta_{y,6\ell a_0/\sqrt{3}} &
    $n$ even \cr 
    N^{-1/2}\delta_{x,na_0} \delta_{y,(6\ell+1) a_0/\sqrt{3}} &
    $n$ odd}, \\
  \phi_{n2}(x,y) & = & \cases{ 
    N^{-1/2}\delta_{x,na_0} \delta_{y,(6\ell-2) a_0/\sqrt{3}} &
    $n$ even \cr 
    N^{-1/2}\delta_{x,na_0} \delta_{y,(6\ell+3) a_0/\sqrt{3}} &
    $n$ odd},
\end{eqnarray}
where the second Kronecker delta function must be satisfied for some
integer $\ell$, and $a_0 = a/2$.  As indicated in Fig.~1, $\phi_{n1}$ and
$\phi_{n2}$ are simply the two normalized basis states with uniform
support at $x=na_0$ on even or odd chains, respectively.  In the
low-energy theory, we may restrict the expansion of the field
operators to this basis:
\begin{equation}
  c^\dagger_\alpha({\bf r}) = \sum_{ni} \phi_{ni}({\bf r})
  c_{ni\alpha}^\dagger.
%  , \qquad c_\alpha({\bf r}) = \sum_{ni} \phi^*_{ni}({\bf r})
%  c_{ni\alpha}. 
\end{equation}
Inserting this into the Hubbard Hamiltonian and summing over
$y$ for fixed $x$ gives
\begin{eqnarray}
  H & = & \sum_n \left\{-t\left( c_{ni\alpha}^\dagger
    c^{\vphantom\dagger}_{n+1,i\alpha} + h.c.\right) - t_\perp \left(
    c_{n1\alpha}^\dagger c^{\vphantom\dagger}_{n2\alpha} +
    h.c.\right)\right\} \nonumber \\ 
& & + u_{N}\sum_{ni} c_{ni\uparrow}^\dagger
c^{\vphantom\dagger}_{ni\uparrow} c_{ni\downarrow}^\dagger
c^{\vphantom\dagger}_{ni\downarrow},
\end{eqnarray}
which is precisely the Hamiltonian of the two-chain Hubbard model, but
with an {\it effective} weak interaction $u_{N} = u/N$.  The factor of
$1/N$ arises because the electrons are delocalized around the
circumference of the nanotube, and hence occupy the same site with a
probability reduced by $1/N$.

A considerable amount is known about the two-chain Hubbard
model,\cite{Hubbard}\ particularly in the weak-coupling limit, where
controlled renormalization group calculations may be used.  These
methods proceed by diagonalizing the kinetic energy, linearizing the
1d spectrum near the resulting Fermi points, and expanding the Hubbard
interactions in the basis of states of the resulting two bands. In the
undoped case at half-filling, interactions drive an instability to a
Mott-insulating spin liquid with a gap in both the charge {\sl and}
spin sectors.  In the weak coupling limit, $u_N \ll t$, both gaps are
exponentially small: $\Delta \sim t \exp(-ct/u_N)$.  At temperatures
below the charge gap, $\Delta_c$, activated behavior is expected in
the resistivity, $\rho \sim \exp(\Delta_c/k_{\rm B}T)$, as illustrated
schematically in Fig.~5.  With increasing $u_N$ the charge gap evolves
continuously into the strong coupling Mott gap associated with the
energy cost of doubly occupying a site.  The spin gap at strong
coupling is more subtle, but indicates a quantum disordered or
short-range resonating valence bond (RVB) ground state.\cite{KRS}\ A
spin gap is also present in a two-leg Heisenberg ladder, in contrast
to a single chain which has gapless spin excitations.\cite{Rice}\ For
the armchair tubes the spin gap at strong coupling can be understood
in terms of a Heisenberg spin model on the honeycomb network, which is
topologically equivalent to the ``brick wall'' lattice shown in
Fig.~4a.  In the anisotropic limit $J_\perp \gg J$, local
spin-singlets form along the vertical rungs, and there is a spin-gap
to triplet excitations.

\begin{figure}[hbt]
\epsfxsize=\columnwidth\epsfbox{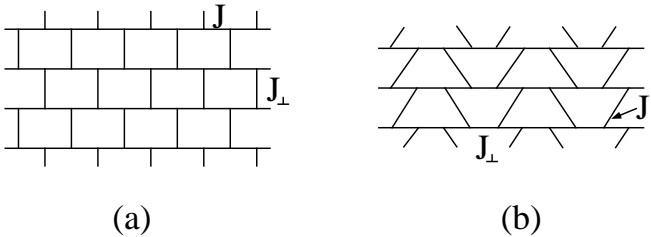}
\vspace{15pt}
\caption{Effective spin models for (a) the armchair tube and (b) the
  zig-zag tube.}
\label{fig4}
\end{figure}

Upon doping for $t_\perp \lesssim 2t$, the two-chain Hubbard model is
known to undergo a phase transition into a state which retains the
spin gap but develops power-law singlet superconducting (SS) and
charge-density-wave (CDW) correlations.\cite{Hubbard}\ Furthermore,
the pair wavefunction associated with the SS correlator has
approximate $d_{x^2-y^2}$ symmetry (i.e. a sign change from quadrant
to quadrant in the $k_x,k_y$ plane).  Both theoretical and numerical
studies\cite{Hubbard,private}\ suggest that the SS correlations are
enhanced over the CDW ones if the Fermi level is pushed into the
proximity of a band edge.  In weak coupling, the enhancement is
mediated by scattering into the nearly empty/full band, for which the
1d van Hove singularity provides an enormous density of states.  This
suggests that for armchair tubes, superconducting effects might be
maximized by tuning the doping so that the Fermi energy coincides with
the lowest-lying ($k_y = \pm 2\pi/Na \sqrt{3}$) bands near the two
Dirac points (see Fig.~3).

For $(N,-N)$ zig-zag tubes with integer $N/3$ the gapless Dirac points
coincide with the discrete quantized momenta for $t = t_\perp$, see
Fig. 2b.  However, due to curvature effects $t \ne t_\perp$, so that
the gapless points are slightly shifted leading to a small
gap,\cite{Theory2}\ of order $|t-t_\perp| \sim 1/N^2$.  Since this is
smaller than the effective interaction strength, which varies as
$1/N$, it is probably legitimate to ignore this small shift when
interactions are included.  Even with this simplification, it is not
possible to map the zig-zag tube directly into a two-chain Hubbard
model.  Nevertheless, proceeding by focusing on the two gapless modes
and expressing the Hubbard interaction in terms of these, one obtains
an effective interacting 1d two-band model for the zig-zag tube.  Just
as for the two-chain Hubbard model, this model has Umklapp scattering
processes, but their particular strengths are different.  It is
natural to expect that these Umklapp processes will gap out both the
charge and spin excitations, just as for the Hubbard model, although a
definitive statement requires a detailed calculation.  Since the $N=3$
zig-zag tube essentially consists of three real-space chains, a spin
gap may seem surprising.  Indeed, it is known\cite{Rice}\ that
conventional Heisenberg spin ladders with an {\it odd} number of legs
have {\it gapless} spin excitations, in contrast to the spin-gapped
even leg ladders.  An important distinction, however, is the unusual
topology of the strong coupling Heisenberg model for the zig-zag tube,
as depicted in Fig.~4b.  In the anisotropic limit, $J_\perp \ll J$,
spins on such a ``herringbone" lattice will indeed form local singlets
across the vertical bonds, with a spin-gap.

Returning to the armchair tubes, since the effective interaction
strength $u_N = u/N$, one expects correlation effects to be weak for
large $N$.  Indeed, as noted above, the gaps in the undoped case
become exponentially small for $u_N \ll t$, and the scale for
superconductivity will likewise be small, indicating the desirability
of reducing the nanotube size in experiments.

Even for larger $N$ (the weak-coupling limit), however, an interesting
observable consequence of interaction physics should remain in the
high-temperature resistivity.  Indeed, Umklapp scattering leads to an
intrinsic contribution to the scattering rate which in weak coupling
varies {\sl linearly} in $T$ for $T \gtrsim \Delta_c$ in
1d!\cite{Millis}\  This is a dramatic enhancement over the conventional
Fermi liquid $T^2$ resistivity, and can also be much larger than
scattering due to (3d) phonons, which vanishes at least as fast as
$T^3$.  We now proceed to obtain a quantitative estimate of this
effect.

The effective low-energy Hamiltonian for the $(N,N)$ nanotube (and for
the two-chain Hubbard model) consists of right and left moving
electrons in the two bands:
\begin{equation}
  H_0 = \sum_{a=1,2} \int dx [ \psi^\dagger_{Ra} iv \partial_x \psi_{Ra}
  - \psi^\dagger_{La} iv \partial_x \psi_{La}] ,
\end{equation}
where we have suppressed the spin label.  Since the equivalent 2-chain
Hubbard model is at half-filling, the presence of the Hubbard
interaction, $u_N$, introduces Umklapp scattering, as well as numerous
momentum conserving four-fermion interactions.  The three
Umklapp interactions, which scatter two right moving electrons into two
left movers, take the form
\begin{equation}
  H_{U} = u_N a_0 \int dx [\psi^\dagger_{L\uparrow} \psi^\dagger_{L
    \downarrow} \psi_{R \downarrow} \psi_{R \uparrow} + h.c. ] ,
\end{equation}
where we have now suppressed the band index.  The scattering rate from
Umklapp scattering can be extracted from the imaginary part of the
electron`s self energy, $\Gamma(\omega,T) = Im \Sigma(\omega,T,k_F)$.
To lowest order there is a single diagram for each of the three
Umklapp interactions, which give identical contributions.  One
finds:
\begin{equation}
  \Gamma(\omega,T) = {3 \over {8 \pi}} (u_N a_0/v)^2 T
  \tilde\Gamma(\omega/2t) , 
\end{equation}
where $\tilde\Gamma(X)$ is a scaling function which approaches $1$ as
$X \rightarrow 0$, and varies as $|X|$ for large $X$.  If we ignore
vertex corrections, the Kubo formula for the 1d
conductivity can be expressed in terms of $\Gamma$ as,
\begin{equation}
  \sigma = {{8v e^2} \over \hbar} \int_{ - \infty}^{\infty} {{d \omega}
    \over {2\pi}} {{(-\partial_\omega f)} \over {\Gamma(\omega,T)}}  ,
\end{equation}
where $f = (e^{\beta \omega} + 1)^{-1}$ is the Fermi function.
The resulting 1d resistivity is
\begin{equation}
  \rho(T) = {c \over 16 \pi} {h \over e^2} (u_N a_0/\hbar v)^2
  (T/\hbar v)  , \label{one_d_res}
\end{equation}
with $c$ a dimensionless constant of order one.

\begin{figure}[hbt]
\epsfxsize=\columnwidth\epsfbox{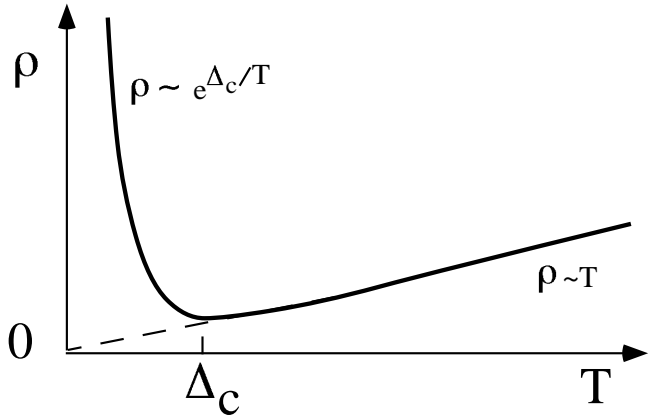}
\vspace{15pt}
\caption{Resistivity of an ideal armchair tube (schematic).}
\label{fig5}
\end{figure}

One of the most promising recent experiments\cite{Ropes}\ studied
single wall carbon nanotubes packed together into a triangular lattice
to form crystalline ``ropes".  These ropes have diameters of 50 to 200
Angstroms, are tens to hundreds of microns long, and are believed to
be predominantly composed of $(10,10)$ armchair tubes.  Transport data
on a single rope reveals a resistivity increasing linearly with
temperature between $50$ and $300K$, consistent with metallic
behavior.  At lower temperatures the resistivity appears to saturate,
perhaps turning up slightly but showing no compelling sign of a
sizeable charge gap.

A comparison with these results can be made by converting
Eq.~(\ref{one_d_res}) to the 3d resistivity $\rho_{3d} \approx \rho
D^2$, where $D$ is the nanotube diameter.  This gives the rough
estimate $\rho_{3d} \sim 2 (u/t)^2 (T/t) \mu \Omega cm$.  Notice that
the nanotube size $N$ has dropped out.  In the experiments, $d
\rho_{exp} / dT \approx 10^{-2} \mu \Omega cm/K$.  To account for this
magnitude one would need a rather large bare Hubbard interaction, $u/t
\sim 10$, perhaps not unreasonable given the neglect of long-ranged
Coulomb forces in our simple Hubbard treatment.  The finite residual
resistivity as $T \rightarrow 0$ is presumably due to disorder.  For
example, local kinks or other defects in the rope packing would
naturally lead to a temperature-independent additive contribution to
the resistivity (see Fig.~5).  However, other effects such as 3d
crossover may also play a role at low temperatures.

We are grateful to Chetan Nayak and Doug Scalapino for helpful
conversations.  This work has been supported by the National Science
Foundation under grants No.  PHY94-07194, DMR94-00142 and DMR95-28578.

\end{document}